\documentclass[fleqn,10pt]{wlscirep}

\usepackage{url}
\usepackage{bbm}
\usepackage{graphicx}
\usepackage{multirow}

\usepackage{color}

\title{An egonet-based approach to effective weighted network comparison}

\author[1,*]{Carlo Piccardi}

\affil[1]{Department of Electronics, Information and Bioengineering, Politecnico di Milano, Piazza Leonardo da Vinci 32, 20133 Milano, Italy}
\affil[*]{carlo.piccardi@polimi.it}

\begin{abstract}
\textbf{With the impressive growth of network models in practically every scientific and technological area, we are often faced with the need to compare graphs, i.e., to quantify their (dis)similarity using appropriate metrics. This is necessary, for example, to identify networks with comparable characteristics or to spot anomalous instants in a time sequence of graphs. While a large number of metrics are available for binary networks, the set of comparison methods capable of handling weighted graphs is much smaller. Yet, the strength of connections is often a key ingredient of the model, and ignoring this information could lead to misleading results. In this paper we introduce a family of dissimilarity measures to compare undirected weighted networks. They fall into the class of alignment-free metrics: as such, they do not require the correspondence of the nodes between the two graphs and can also compare networks of different sizes. In short, they are based on the distributions, on the graph, of a few egonet features which are easily defined and computed: the distance between two graphs is then the distance between the corresponding distributions. On a properly defined testbed with a pool of weighted network models with diversified characteristics, the proposed metrics are shown to achieve state-of-the-art performance in the model classification task. The effectiveness and applicability of the proposed metrics are then demonstrated on two examples. In the first, some ''filtering'' schemes -- designed to eliminate non-significant links while maintaining most of the total weight -- are evaluated in their ability to produce as output a graph faithful to the original, in terms of the local structure around nodes. In the second example, analyzing a timeline of stock market correlation graphs highlights anomalies associated with periods of financial instability.}
\end{abstract}

\begin{document}

\flushbottom
\maketitle

\thispagestyle{empty}



Networks are modeling tools universally adopted to describe the backbone of interactions among the multiple components of a complex system \cite{Ne:10,Ba:16,Latora2017}. However, not all these interactions are equivalent: in the most varied fields of application, it has been recognized that different interactions are often associated with different degrees of intensity – hence the need to provide each edge of the network with a weight that quantitatively expresses that intensity. To cite just a few of the most studied examples, weights can represent the strength of ties between individuals in social networks\cite{Granovetter1973}, the quantity of cooperative outcomes in a scientific collaboration network\cite{Newman2001}, fluxes in metabolic reactions\cite{Almaas2004}, the counting of passengers in air connection networks\cite{Colizza2006}, the value of goods exchanged between countries (or companies) in trade networks\cite{SeBo:07,FaRe:08,servicenet2021}, and the level of synchrony of time series signals in correlation networks, which find application, for example, in the modeling of stock markets\cite{Mantegna1999,Onnela2004}, large-scale climate systems\cite{Donges2009}, or brain networks\cite{RUBINOV2010}. In all these cases, neglecting the weights, i.e., trivially adopting a binary interaction structure, would imply a crucial loss of information and lead to an incomplete (and perhaps misleading) analysis of the properties of the system. The limit case is correlation networks, which are generally cliques (all-to-all) by construction, so that their study must rely completely on weights. But even when the graph is sufficiently sparse, there are cases -- for example the world trade network -- where connections carry weights that continuously span many orders of magnitude. In this case the analysis of the network of pure interactions would lead to a completely distorted picture, for example in terms of which agents are most central or which connections are most strategic. The importance of analyzing networks taking full account of the intensity of connections has given rise, in the last two decades, to a vast literature of theories and methods suitable for the study of what are called \textit{weighted networks} (e.g., Refs.\cite{Barrat2004,Ne:04,Onnela2005,Thada2005,Serrano2006,DaBa:06,Latora2017}).

A problem that has received, up to now, only limited attention in the study of weighted networks is that of \textit{comparison}, that is, the quantification of similarities or differences between weighted graphs. This is a fundamental problem that arises when analyzing a set of networks, for example with the aim of finding groups of systems with similar characteristics, or of detecting  discontinuities in the time pattern when a set of temporally ordered networks is given. If we focus on binary (i.e., unweighted) network models, we find a large amount of literature discussing applications of comparison techniques to diverse fields (e.g., Refs.\cite{Pr07,van:10,Ali:14,SoEl:14,PiPi:20,brain:20,servicenet2021}). A very large number of methods and approaches are available, with different levels of applicability and effectiveness: for surveys we refer to Refs.\cite{EmDe:16,DoHo:18,Ta19}. On the other hand, only very few methods are available to effectively compare weighted networks, especially if we refer to the general family of alignment-free methods, which try to capture the difference in the global structure rather than the discrepancies in the neighborhood of each node. These methods do not require matching between nodes, so that a measure of (dis)similarity can be defined between any pair of graphs, even with different sizes and densities.

In this paper, we propose an alignment-free approach to define a set of dissimilarity measures between two (undirected) weighted graphs. This is an improvement and generalization of a method for binary networks recently introduced\cite{Pi23} which, on a standard benchmark pool of network models, achieves performance comparable to those of the graphlet-based measures\cite{Ya2014,Ya2015} and, therefore, it lends itself to being candidate for high quality performance even in the most demanding context of weighted networks. In the proposed approach, each network is summarized by the distribution of three indicators that locally describe the 1-step egonet of each node: the weighted degree (or strength) of the node, the clustering coefficient, and the egonet persistence. The three indicators, which are standard in network science and can be calculated rather quickly, not only capture the local topological structure around the ego-node, but also fully take into account the weights of all relevant connections, as will be clear by their definition (see \textit{Methods}). They describe in increasingly more detail the connectivity inside the egonet and across its border, and can be used to define 1- to 3-dimensional distribution functions that are taken as summaries of the network properties. The dissimilarity between two graphs is then defined as the distance between the corresponding distributions.

In the literature, a baseline method for defining the dissimilarity between two weighted graphs is to calculate the difference between one or more global (scalar) quantities, such as the clustering coefficient, average shortest path, or diameter. This approach, however, has been proven to have limited success on binary networks \cite{Pr07,Ta19} and there is no reason to believe that it should be more effective in an even more challenging context. For benchmark purposes, we will evaluate the performance of using the global (weighted) \textit{clustering coefficient} as a dissimilarity metric. In the same spirit we will also compare our proposal to \textit{spectral metrics}, which are based on the comparison of the set of eigenvalues of the (weighted) adjacency or Laplacian matrices\cite{Wilson2008}. In the set of benchmark methods we will also evaluate {\textit{WD-metric}} and \textit{Portrait Divergence}. {The former is defined as a linear combination of three distance terms\cite{Jiang2021,Schieber2017}: two of them compare, with appropriate metrics, the distributions of the (weighted) shortest-path lengths in the two graphs, while the third compares the distributions of the alpha-centralities in the two graphs and their complements.} Portrait Divergence, on the other hand, defines a distance by comparing the ''portrait'' of the two graphs, i.e., a matrix that encodes the distribution of the lengths of the shortest-path\cite{Bagrow2019}, and behaves very well in classifying binary networks\cite{Ta19}. It has been generalized to define a measure of dissimilarity between weighted networks (Ref.\cite{Bagrow2019}, Supplemental Material), although it has not been extensively tested in this regard. Finally, we note that the class of graphlet-based alignment-free methods \cite{Ya2014,Ya2015}, which guarantees very high-level performance on binary networks \cite{Ta19,Pi23}, does not yet have a generalization to weighted networks, as discussed in Ref.\cite{Newaz2020}.

When evaluating methods for comparing weighted networks, a major issue is the lack of an adequate testbed that generalizes the benchmark pool of models used for binary networks\cite{Ya2014,Ya2015}. A contribution of this paper is to first define such a testbed by introducing a set that includes 12 weighted networks models: 9 of them are obtained from 3 well-known binary models (Erd\H{o}s-R\'{e}nyi, Barab\'asi-Albert, and Geometric Random Graphs) by imposing 3 different weighting schemes on each. The remaining ones (Yook-Jeong-Barab\'asi-Tu model\cite{Yook2001}, and Antal-Krapisvky model with random or exponential weighting\cite{Antal2005}) were originally proposed for the construction of weighted networks. As will be highlighted, the testbed contains sufficiently diverse models to explore a broad spectrum of graph characteristics and, at the same time, models that differ only in subtle features, in order to intentionally challenge the methods under test.

The classification capabilities of the proposed ''weighted ego-distances'' will be extensively evaluated on the testbed defined above. More precisely, the aim is to discriminate pairs of graphs originating from the same model from those originating from different models, with size and density (i.e., number of nodes and edges) acting as confounding factors. Note that the definition of ''model'' includes the weighting scheme, so, for example, a dissimilarity measure should be able to recognize that two Barab\'asi-Albert networks have different weighting schemes, even if they have exactly the same topology. The results of the experiments show that the proposed weighted ego-distances are able to outperform all distances used for comparison, and therefore should be considered, to date, the most advanced method for comparing undirected weighted networks. {This statement is, of course, based on the set of benchmark measures and the composition of the network model pool used in the experimental setup: both, however, have been defined to the best of our knowledge.}

The paper concludes with two application examples. In the first, a weighted ego-distance is used to compare the effects of different algorithms to filter weighted networks, i.e., to derive a ''backbone'' by eliminating a large number of less significant edges while maintaining most of the total network weight\cite{Yassin2023}. Three filtering schemes are evaluated, namely the trivial hard threshold, the classical disparity filter\cite{SeBo:09}, and the more recent P\'{o}lya filter\cite{Marcaccioli2019}. Their application to two different datasets (the world trade network and the US airport network) reveals the superiority of the P\'{o}lya filter, in the sense that, at a given level of pruning, it is able to preserve the egonet (i.e., local) features of the original graph to a greater extent. The second application example concerns a temporal sequence of correlation networks, each obtained by analyzing the relationships between 3-month long time series referring to the stocks that define the S\&P100 index. The pairwise comparison between the time-indexed correlation networks confirms a well-known result, namely that, given the greater coherence between the time series, the correlation graphs associated with financial crises have significantly different characteristics from those characterizing ordinary periods\cite{ONNELA2003,Onnela2005,Lacasa2015}. 


\section*{Results}

\subsection*{Ego-distances}

\begin{figure}
	\centering
	\includegraphics[width=16.5cm]{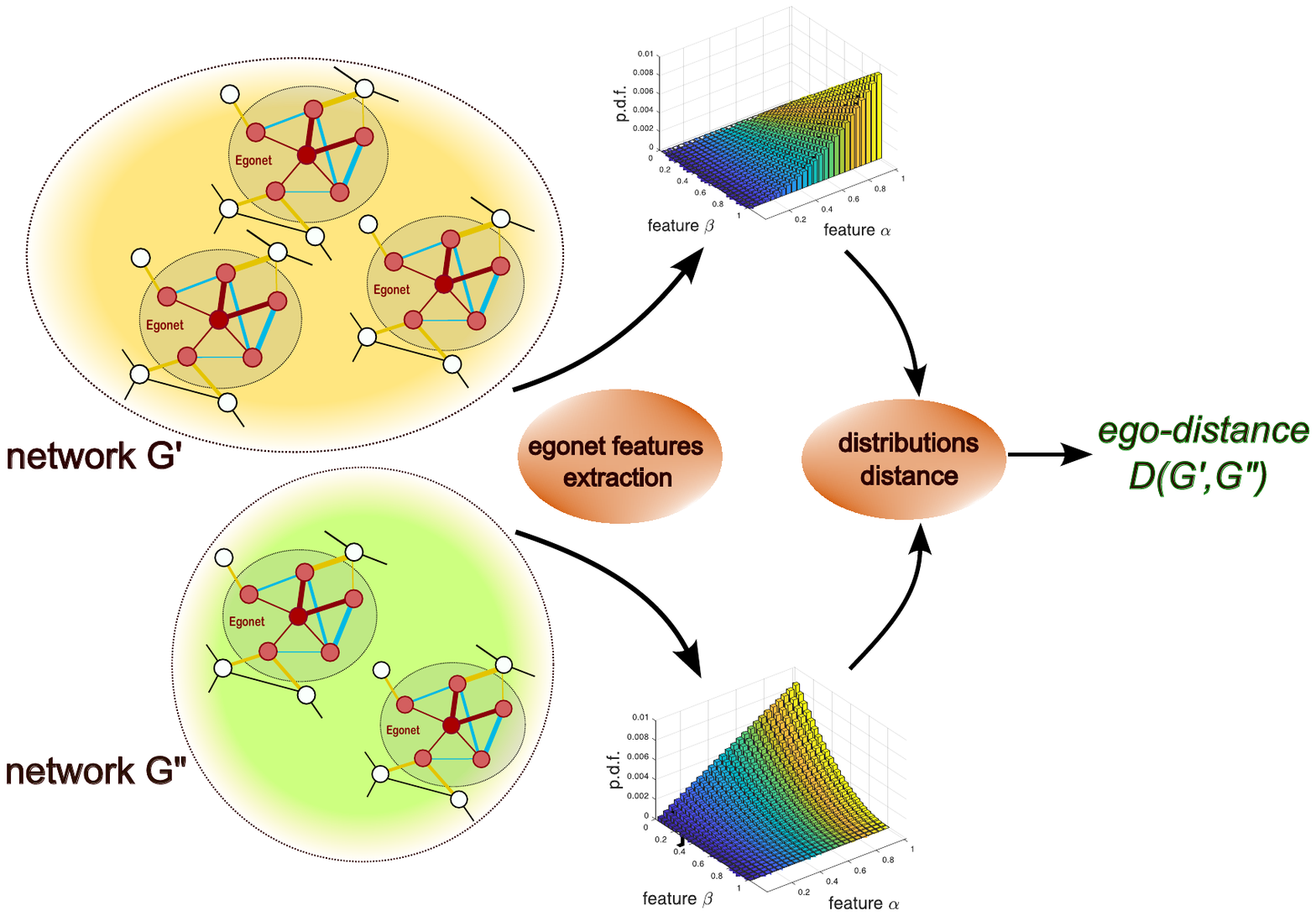}
	\caption{\label{fig:summary} A summary of the workflow for calculating the ego-distance $D(G',G'')$. For each undirected weighted graph, all egonets are analyzed to calculate some features, which are then exploited to define from 1- to 3-dimensional distribution functions (the case of 2-dimensional distributions is illustrated in the figure as an example). The graph-to-graph distance $D(G',G'')$ is then defined as the distance between the two distributions. }
\end{figure}

Figure \ref{fig:summary} summarizes the workflow of the operations that define the ego-distance $D(G',G'')$. We are given the two undirected weighted graphs $G'$, $G''$, in general of different sizes $N_1$, $N_2$ and densities $\rho_1$, $\rho_2$. The egonet of each node (i.e., the subgraph including the node and its nearest neighbors) is analyzed and some features are calculated: the (normalized) \textit{weighted degree} (or strength) $d$, the \textit{clustering coefficient} $c$, and the \textit{egonet persistence} $p$ (see \textit{Methods}). They characterize the local connectivity around the ego-node, both in terms of structure (topology) and intensity (weights). The egonet features values on each network are then used to define single- or multi-feature distribution functions, one for each network, and the distance between the two distributions is taken as the distance between the two graphs. More precisely, we define the following ego-distances:
\begin{itemize}
	
	\item $D_d$, $D_c$, $D_p$ (1-feature distances): The distance between the distributions of $d$, $c$, or $p$, respectively.
	
	\item $D_{cp}$, $D_{dc}$, $D_{dp}$ (2-feature distances): The distance between the respective 2-dimensional distributions (this is the case exemplified pictorially  in Fig. \ref{fig:summary}).
	
	\item $D_{dcp}$ (3-feature distance): The distance between the 3-dimensional distributions.
	
	\item $D_{SUM}$ (3-feature distance): It is proportional to the sum of the 1-feature distances $D_d$, $D_c$, $D_p$ defined above. It uses all the features available, as $D_{dcp}$ does, but in a simplified form that makes it computationally more feasible.		
	
\end{itemize}

\subsection*{\label{sec:net_models}Weighted network models}

We evaluate the effectiveness of the ego-distances proposed above by generalizing the experimental setup defined in Refs. \cite{Ya2014,Ya2015,Pi23} with reference to unweighted networks. It is necessary to extend the above framework to the case of weighted networks for which, to the best of our knowledge, a standard benchmark pool of network models to be used for classification has never been defined to date. 

To this end, we consider a set of $12$ undirected weighted network models, as described below. For each model, we generate networks with three different sizes (number of nodes) and three different densities, for a total of $12\times 3\times 3=108$  model/size/density combinations. For each combination, we randomly generate $10$ network instances, so that the experimental setup includes $1080$ networks. The $12$ network models are as follows (see \textit{Methods} for details):

\begin{itemize}
	
	\item \textbf{ER-x}: \textit{Erd\H{o}s-R\'{e}nyi} model with \textit{Uniform} (ER-U), \textit{Random} (ER-R), or \textit{Degree-dependent} (ER-D) weighting.
	
	\item \textbf{BA-x}: \textit{Barab\'asi-Albert} model with \textit{Uniform} (BA-U), \textit{Random} (BA-R), or \textit{Degree-dependent} (BA-D) weighting.
	
	\item \textbf{GEO-x}: \textit{geometric random graph} model with \textit{Uniform} (GEO-U), \textit{Random} (GEO-R), or \textit{Degree-dependent} (GEO-D) weighting.
	
	\item \textbf{YJBT}: \textit{Yook-Jeong-Barab\'asi-Tu} model\cite{Yook2001}.
	
	\item \textbf{AK-x}: \textit{Antal-Krapisvky} model with \textit{Random} (AK-R) or \textit{Exponential} (AK-E) weighting\cite{Antal2005}.
	
\end{itemize}

\begin{figure}
	\centering
	\includegraphics[width=12.5cm]{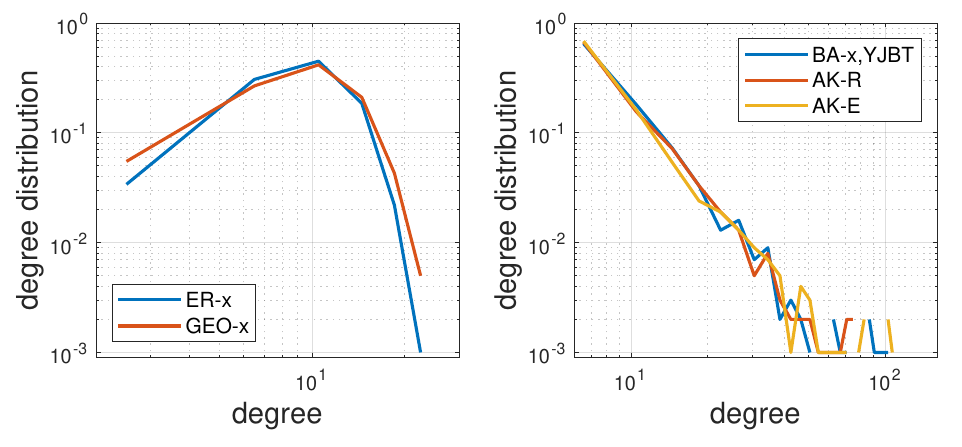}
	\caption{\label{fig:deg-distributions} Examples of degree distribution for the $12$ synthetic network models   ($N=1000$, $\rho=0.01$). By construction, the three ER-x models (with different weighting schemes) have the same degree distribution; the same for the three GEO-x models, and for the three BA-x models together with the YJBT model.}
\end{figure}

The above network models can be neatly divided into two main sets based on their degree distribution, as illustrated in Fig. \ref{fig:deg-distributions}. ER-x and GEO-x have single-scale (Poisson) distribution (we will refer to them as \textit{degree-homogeneous networks}), while the others (BA-x, YJBT, AK-x) have scale-free (power-law) distribution\cite{BoLa:06} (\textit{degree-heterogeneous networks}).

\begin{figure}
	\centering
	\includegraphics[width=17.5cm]{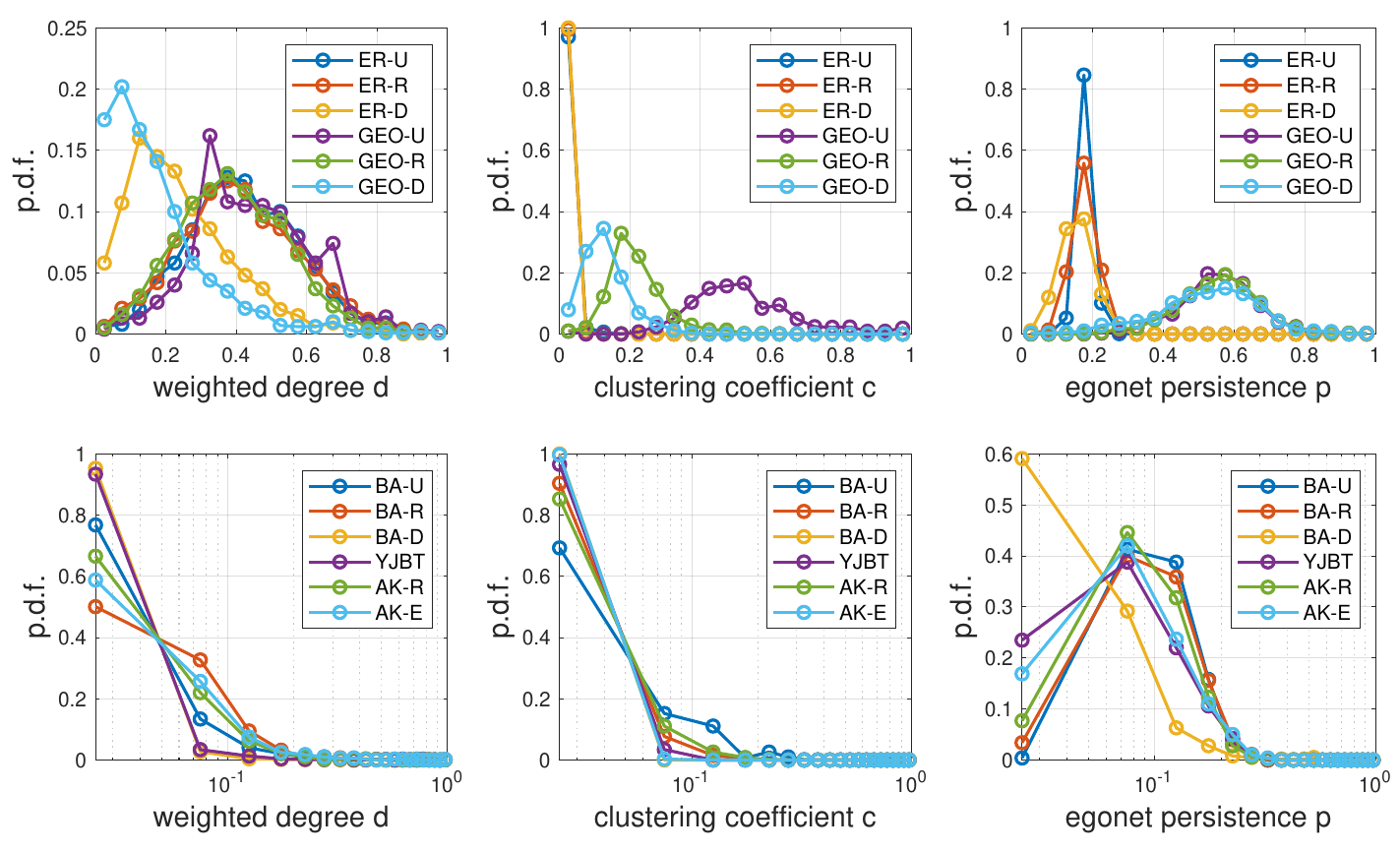}
	\caption{\label{fig:distributions} Example distributions of (normalized) weighted degree (left column), clustering coefficient (middle column), and egonet persistence (right column) for the $12$ synthetic network models ($N=1000$, $\rho=0.01$). Degree-homogeneous networks are in the top row, degree-heterogeneous networks in the bottom row {(where a logarithmic x-axis is used to improve visibility)}.}
\end{figure}

The picture becomes more complex when different weighting schemes are considered and, consequently, when egonet features are calculated on the weighted networks. Figure \ref{fig:distributions} shows examples of egonet feature distributions. The top row, which refers to degree-homogeneous networks, shows that ER-x are quite well recognizable from GEO-x by the clustering coefficient (essentially zero in ER-x, for all weighting schemes) and by the egonet persistence (on average much higher in GEO-x due to clustering). Within the two families ER-x and GEO-x, the weighting schemes give rise to different distributions of the clustering coefficient in GEO-x, and of the egonet persistence in ER-x (although with less differentiation). Furthermore, degree-dependent weighting stretches the distribution of the weighted degree. Taken together, the six models ER-U/R/D and GEO-U/R/D appear quite recognizable from each other, and indeed we anticipate that model classification works very well, if limited to these models (see \textit{Model classification}). 

If instead we consider the bottom row of Fig. \ref{fig:distributions}, we notice that the strong heterogeneity in the degree induces similar heterogeneity in at least two of the egonet features, namely weighted degree and clustering coefficient. Here the classification task appears much more difficult but, as we will see, the proposed ego-distances manage to exploit the subtle quantitative differences between the various distributions and give a good performance overall. It goes without saying that, compared to the mere visual analysis of the Fig. \ref{fig:distributions} outlined here, the classification task is made more challenging by the different sizes and densities of the graphs compared.

\subsection*{Model classification}

For each pair of networks, we calculate the eight ego-distances defined above (see \textit{Ego-distances}). To have a comparison in terms of classification performance, we also calculate the following benchmark distances\cite{Ta19} (see \textit{Methods} for details):
\begin{itemize}
	
	\item $D_{\text{C}_{global}}$: The distance between two graphs is the difference of their average (weighted) clustering coefficients. Although very simple, this metrics has proven to be rather effective for unweighted networks \cite{Ya2014,Ya2015,Pi23} while, to  our knowledge, it has never been tested in the weighted case.
	
	\item $D_{SP-W/L}$: Spectral distance based on the \textit{Weight} (SP-W) or \textit{Laplacian} (SP-L) matrix. They are defined as the distance between the spectra of the respective graph matrices\cite{Wilson2008}. 
	
	\item {$D_{WD}$ (WD-metric): It is based on a combined indicator that compares the distributions of the (weighted) shortest-path lengths of the two graphs, as well as the distributions of the alpha-centralities in the two graphs and their complements\cite{Jiang2021,Schieber2017}.}
	
	\item $D_{PDiv}$ (Portrait Divergence): It is based on the comparison of the \textit{portrait matrices} of the two networks, which encode the distribution of the (weighted) shortest-path lengths of the graphs\cite{Bagrow2019}. 
	
\end{itemize} 

The classification experiments follow the pairwise comparison framework adopted in Refs.\cite{Ya2014,Ya2015,Pi23}. The task is to recognize when two networks have been generated by the same model and, in this regard, a measure is effective if the distance between two graphs generated by the same model is significantly smaller than the distance between two graphs coming from different models. Non-coinciding size and/or density in the two graphs act as confounding factors, as they could hide possible structural differences. To account for this factor, we will evaluate the performance of each distance separately on both the entire pool of networks and  the subset of network pairs having the same size and density. The performance of each distance is evaluated within the usual Precision-Recall (PR) framework, which quantifies its ability to correctly classify if a network pair is generated by the same model (see \textit{Methods}): the quantity AUPR (\textit{Area Under the Precision-Recall curve}, $0\le \text{AUPR}\le 1$) summarizes the performance of each distance, with the limit $\text{AUPR}=1$ obtained in the ideal case\cite{Davis2006,Saito2015}.

Table \ref{tab:all_networks} summarizes, in terms of AUPR value, the results of the classification experiments described above. To combine the performance obtained by comparing graphs with different and with the same size/density, we also included the average of the two AUPR values (last column) as an overall indicator. In general, all ego-distances that use a combination of egonet features (i.e.,  $D_{SUM}$, $D_{cp}$, $D_{dc}$, $D_{dp}$, and $D_{dcp}$) perform better than the benchmark distances ($D_{C_{global}}, D_{SP-W/L}, {D_{WD}}, D_{PDiv}$) in most cases. {Partial exceptions are $D_{WD}$ (WD-metric) and $D_{PDiv}$ (Portrait Divergence), which perform similarly to ego-distances when comparing graphs of the same size/density, although their performance drops significantly more than ego-distances when considering mixed size/density.} Overall, $D_{dcp}$, which exploits all the information available in the most extensive form, obtains the best performance on average. A viable and computationally simpler alternative is $D_{SUM}$, which achieves only slightly worse performance but is computationally lighter. {Incidentally, we note that there is no clear evidence that one of the egonet features is less important than the other, to the point that it could be dropped: for example, $D_c$ performs the worst among the three 1-feature distances (see the last column \textit{avg}), suggesting that the clustering coefficient might be less relevant, yet $D_{dc}$ -- which does exploit the clustering coefficient -- is the best among the 2-feature distances.} 

\begin{figure}
	\centering
	\includegraphics[scale=.6]{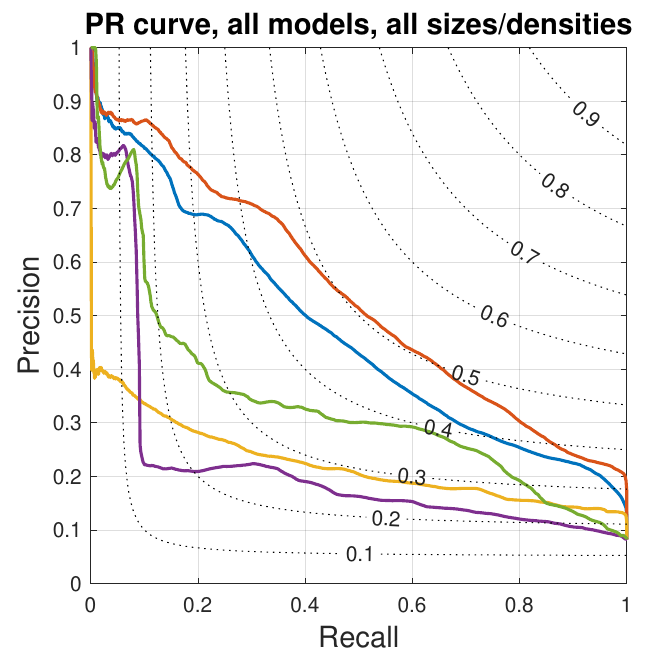}
	\includegraphics[scale=.6]{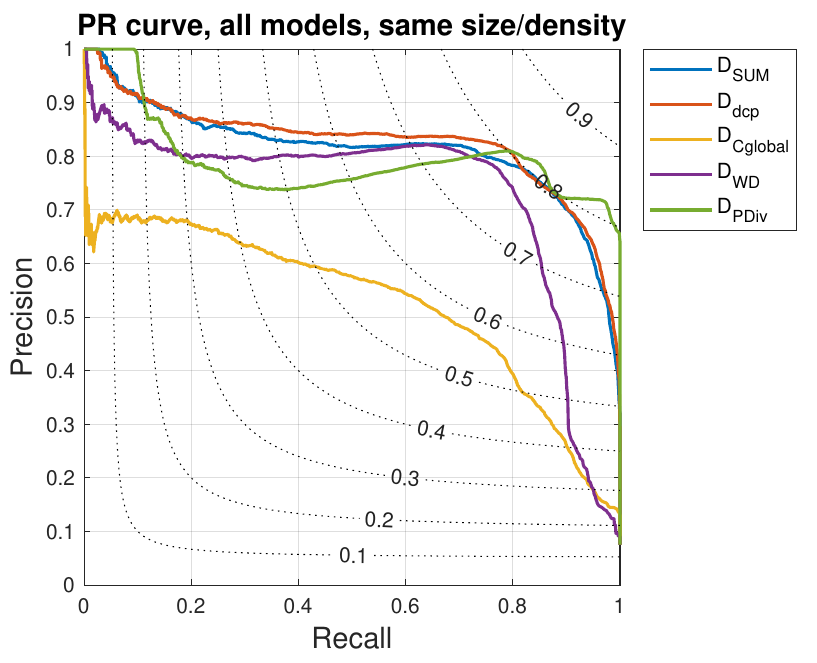}
	\caption{\label{fig:PR} Precision-Recall curves obtained by classifying the complete set of networks (12 models $\times$ 3 sizes $\times$ 3 densities $\times$ 10 replications = 1080 networks) with two of the ego-distances defined in the section \textit{Methods}  ($D_{SUM}$, $D_{dcp}$) and {three} of the benchmark distances ($D_{C_{global}}$, {$D_{WD}$}, $D_{PDiv}$) (the curves related to all other distances are omitted for readability). For each Recall-Precision point, the F1 value (i.e., the harmonic mean of Precision and Recall) is specified by the dotted contour line. Left: Performance is assessed by mixing sizes/densities; Right: Only networks with the same size/density are compared. }
\end{figure}

The Precision-Recall curves of Fig. \ref{fig:PR} (left panel) show that the ego-distances which exploit all the available information ($D_{dcp}$ and $D_{SUM}$) achieve similar behavior to each other, and outperform the benchmark distances ($D_{PDiv}$ and to a larger extent $D_{C_{global}}$ {and $D_{WD}$}). $D_{dcp}$ provides the best trade-off at high Recall values, as evidenced by the largest F1 value slightly above $0.5$ (we recall that F1 is the harmonic mean of Precision and Recall). 
On the other hand, when the comparison is limited to networks with the same size/density, Fig. \ref{fig:PR} (right panel) confirms that, as underlined in Table \ref{tab:all_networks}, $D_{dcp}$, $D_{SUM}$, and $D_{PDiv}$ achieve similar performance, {slightly better than $D_{WD}$ at high Recall values,} while $D_{C_{global}}$ follows with a large gap {(further experiments on the dependence of the results on size and density are in the \textit{Supplementary Information} file). }

\begin{table}
	\centering
	\caption{\label{tab:all_networks} AUPR (Area Under the Precision-Recall curve) value for the classification of all network models, for the ego-distances (top part of the table) and for the benchmark distances (bottom part). The last column contains the average of the previous two. The best value in each column is in bold.}
	\begin{tabular}{cccc}
		\toprule
		{distance}&{all sizes/densities}&{same size/density}&\textit{avg}\\
		$D_d$ & 0.352 & 0.494 & \textit{0.423} \\
		$D_c$ & 0.210 & 0.440 & \textit{0.325} \\
		$D_p$ & 0.265 & 0.672 & \textit{0.469} \\
		$D_{SUM}$ & 0.471 & 0.814 & \textit{0.642} \\
		$D_{cp}$ & 0.370 & \textbf{0.864} & \textit{0.617} \\
		$D_{dc}$ & 0.503 & 0.745 & \textit{0.624} \\
		$D_{dp}$ & 0.410 & 0.696 & \textit{0.553} \\
		$D_{dcp}$ & \textbf{0.536} & 0.826 & \textit{\textbf{0.681}} \\
		\midrule
		$D_{\text{C}_{global}}$ & 0.223 & 0.527 & \textit{0.375} \\			
		$D_{SP-W}$ & 0.224 & 0.679 & \textit{0.452} \\
		$D_{SP-L}$ & 0.195 & 0.568 & \textit{0.381} \\
		{$D_{WD}$}   & {0.219} & {0.729} & {\textit{0.474}} \\ 
		$D_{PDiv}$ & 0.334 & 0.794 & \textit{0.564} \\	
		\bottomrule		
		\multicolumn{4}{l}{{\footnotesize{$D_{d},D_c,D_p$: 1-feature ego-distances; $D_{cp},D_{dc},D_{dp}$: 2-feature ego-distances; }}} \\
		\multicolumn{4}{l}{{\footnotesize{$D_{SUM},D_{dcp}$: 3-feature ego-distances; $D_{\text{C}_{global}}$: global clustering coefficient;}}} \\
		\multicolumn{4}{l}{{\footnotesize{$D_{SP-W}$: spectral, weight matrix; $D_{SP-L}$: spectral, Laplacian matrix;}}} \\
		\multicolumn{4}{l}{{\footnotesize{$D_{WD}$: WD-metric; $D_{PDiv}$: Portrait Divergence }}} \\
	\end{tabular}
\end{table}

To delve deeper into the behavior of the analyzed network distances, we reproduce the experiments summarized in Table \ref{tab:all_networks} after having decomposed the pool of networks in two subsets, respectively containing the $6$ models with homogeneous degree distribution (ER-U/-R/-D, GEO-U/-R/-D) and the $6$ models with heterogeneous degree distribution (BA-U/-R/-D, YJBT, AK-R/-E), as highlighted in Fig. \ref{fig:deg-distributions}. The results are in Table \ref{tab:breakdown}. All distances perform well in classifying degree-homogeneous networks: ER and GEO networks are topologically different (the former has vanishing clustering coefficient, unlike the latter) and, within the two models, the metrics analyzed recognize the different weighting schemes quite easily (Fig. \ref{fig:distributions}). Classification is almost perfect when comparing graph with the same size/density, while it is more problematic with different sizes/densities, and ego-distances are here more effective. More challenging is the classification in the pool of degree-heterogeneous networks, whose features present much more subtle differences (Figs. \ref{fig:deg-distributions} and \ref{fig:distributions}). Nonetheless, ego-distances still guarantee good performance, which becomes very good in the case of same size/density.

\begin{table}
	\centering
	\caption{\label{tab:breakdown} AUPR (Area Under the Precision/Recall curve) value for the classification of network models, for the ego-distances (top part of the table) and for the benchmark distances (bottom part) (we omit the distances $D_d$, $D_c$, $D_p$, and $D_{SP-W/L}$, whose performance is lower in most cases). The classification is performed separately for the network models with homogeneous or heterogeneous degree distribution. The best value in each column is in bold. }
	\begin{tabular}{c|ccc|ccc}
		\multicolumn{1}{c}{} & \multicolumn{3}{|c|}{degree-homogeneous networks} & \multicolumn{3}{|c}{degree-heterogeneous networks} \\
		\multicolumn{1}{c}{} & \multicolumn{3}{|c|}{ER-U/-R/-D, GEO-U/-R/-D} & \multicolumn{3}{|c}{BA-U/-R/-D, YJBT, AK-R/-E}\\
		\toprule
		{distance}&{all sizes/densities}&{same size/density}&\textit{avg}&{all sizes/densities}&{same size/density}&\textit{avg}\\
		$D_{SUM}$ & 0.680 & 0.949 & \textit{0.814}& 0.470 & \textbf{0.845} & \textit{0.657} \\
		$D_{cp}$ & 0.632 & {0.980} & \textit{0.806}& 0.340 & {0.802} & \textit{0.571} \\
		$D_{dc}$ & 0.628 & 0.876 & \textit{0.752} & \textbf{0.540} & 0.816 & \textit{\textbf{0.678}}\\
		$D_{dp}$ & 0.547 & 0.723 & \textit{0.635} & 0.365 & 0.774 & \textit{0.569} \\
		$D_{dcp}$ & {\textbf{0.749}} & 0.935 & \textit{{\textbf{0.842}}}& {0.445} & 0.843 & \textit{{0.644}}  \\
		\midrule
		$D_{\text{C}_{global}}$ & 0.456 & 0.679 & \textit{0.568}& 0.386 & 0.674 & \textit{0.530} \\		
		{$D_{WD}$} & {0.360} & {0.876} & {\textit{0.618}} & {0.278} & {0.677} & {\textit{0.478}} \\	
		$D_{PDiv}$ & 0.547 & \textbf{0.999} & \textit{0.773} & 0.516 & 0.611 & \textit{0.563}\\	
		\bottomrule		
	\end{tabular}
\end{table}

{A complementary analysis to the above is to group the network models based on their weighting scheme, rather than their degree distribution properties. This defines three groups, respectively with uniform  (ER-/BA-/GEO-U), random (ER-/BA-/GEO-R, AK-R), or degree-dependent weighting (ER-/BA-/GEO-D, YJBT) -- note that AK-E (with exponential weighting) is excluded because it has no other models to compare to. The results are in Table \ref{tab:breakdown2}. Models with uniform weights (namely binary networks) are classified almost perfectly by ego-distances, despite mixed size/density acting as a confounder. Performance deteriorates slightly when considering random and degree-dependent weightings, although ego-distances remain significantly better. It is worth mentioning that much of the performance degradation reported in the second column (random weighting) is due to the inclusion of the AK-R model, which is only slightly different from BA-R (see Methods for details on the definition of the two models). If we remove AK-R from the pool and leave only ER-/BA-/GEO-R, the AUPR increases sharply for all measures (with the benchmark measures still dominated by ego-distances) reaching, e.g., $D_{dcp}=0.999$ or $D_{SUM}=0.993$. The same result occurs when removing YJBT from the third pool, given its similarity to BA-D (the two models have exactly the same topology and a slightly different weighting rule). Overall, this confirms that when considering a pool of networks with similar weighting scheme, the topological structure (not surprisingly) plays a crucial role for effective classification.  }

\begin{table}
	\centering
	\caption{\label{tab:breakdown2} {AUPR (Area Under the Precision/Recall curve) value for the classification of network models, for the ego-distances (top part of the table) and for the benchmark distances (bottom part) (we omit the distances $D_d$, $D_c$, $D_p$, and $D_{SP-W/L}$, whose performance is lower in most cases). The classification is performed separately for the network models with uniform, random, or degree-dependent weighting. All sizes/densities are included in the experiment. The best value in each column is in bold.} }
	{
		\begin{tabular}{c|ccc}
		\multicolumn{1}{c}{} & \multicolumn{1}{|c}{uniform weighting} & \multicolumn{1}{c}{random weighting} & \multicolumn{1}{c}{degree-dependent weighting} \\
		\multicolumn{1}{c}{distance} & \multicolumn{1}{|c}{ER-/BA-/GEO-U} & \multicolumn{1}{c}{ER-/BA-/GEO-R, AK-R} & \multicolumn{1}{c}{ER-/BA-/GEO-D, YJBT} \\
		\toprule
		$D_{SUM}$ & 0.993 & 0.639 & 0.686 \\
		$D_{cp}$ & 0.827 & 0.559 & 0.643 \\
		$D_{dc}$ & \textbf{0.999} & 0.631 & 0.669 \\
		$D_{dp}$ & 0.993 & 0.664 & 0.704 \\
		$D_{dcp}$ & \textbf{0.999} & \textbf{0.683} & \textbf{0.730} \\
		\midrule
		$D_{\text{C}_{global}}$ & 0.774 & 0.550 & 0.463 \\		
		$D_{WD}$ & 0.581 & 0.408 & 0.567 \\	
		$D_{PDiv}$ & 0.515 & 0.348 & 0.544 \\	
		\bottomrule		
	\end{tabular} }
\end{table}

{To summarize the results reported above, and to provide guidance on which of the proposed ego-distances to adopt, we firstly note that AUPR should be the primary selection criterion, as it represents the average precision over all possible recall values. When all network models are included in the classification problem (Table \ref{tab:all_networks}), it is not surprising that AUPR highlights the superiority of measures that exploit all available information: $D_{dcp}$ and $D_{SUM}$ have the best average performances, and the latter is a valid alternative to the former having a lower computational load (see \textit{Methods}). Note that, in the ''all sizes/density'' case, $D_{dcp}$ remains the best measure, with $D_{SUM}$ also achieving a very good performance, while in the ''same size/density'' case the two measures, although not at the top, follow with a very small gap. The same excellent behavior of these two measures consistently emerges from the experiments in Tables \ref{tab:breakdown} and \ref{tab:breakdown2}, where the network models are split based on their degree distribution or weighting characteristics, respectively. In all cases, $D_{dcp}$ and $D_{SUM}$ provide the best performance or, in the worst case, a value close to the optimal one.}

\subsection*{\label{sec:filters}Example of application: Filtering multiscale weighted networks}

We demonstrate the use of the ego-distances introduced above by comparing the effect of different filters applied to weighted networks. Numerous methods have been proposed in the literature to filter weighted networks\cite{Yassin2023}, i.e., to obtain a simplified network (a so-called ''backbone'') that maintains most of the global weight of the original one while pruning a large number of insignificant connections. The problem is relevant and non-trivial in multiscale networks, where link weights span many orders of magnitude and therefore nodes have very heterogeneous features. Here, the trivial ''hard filtering'' scheme, which consists in removing all weights below a given threshold, is not suitable because it cannot handle the multiscale nature of the graph -- for example, it could systematically isolate the nodes with small strength.

Our exercise consists in evaluating the trend of the distance between a multiscale weighted network and its filtered versions while varying the selectivity parameter of the filter, i.e., moving from the original network to increasingly pruned graphs. We rely on the ego distance $D_{dcp}$ defined above, and use three different filtering schemes, each characterized by a single parameter:
\begin{itemize}
	\item \textbf{Hard threshold}: Once a value $0<\gamma<1$ is set, all links with weight $w_{ij}<\gamma w_{\max}$ are discarded ($w_{\max}$ is the largest weight in the network).
	\item \textbf{Disparity filter}\cite{SeBo:09}: It is based on a null model that assumes a uniformly random distribution of a node's strength across its connections. Links that deviate from this assumption are significant, according to a prescribed $\alpha$ level, and therefore retained in the filtered network.
	\item \textbf{P\'{o}lya filter} \cite{Marcaccioli2019}: Here the null model assumes that a node distributes its strength among its links following a P\'{o}lya process, rather than uniformly, with reinforcement governed by a parameter $a>0$.
\end{itemize}

We consider two networks. In the \textit{World Trade Network} (WTN\cite{SeBo:07,FaRe:08,Hoang2023}) the nodes are the countries ($N=223$) and the weight of the link $i\rightarrow j$ is the total value (in USD) of goods of any category sold from country $i$ to country $j$ in a given year (we use 2014 data). We symmetrize the graph, so the weight $w_{ij}=w_{ji}$ of the link $(i,j)$ is the  total bilateral trade between the two countries. Weights and strengths are extremely different, covering eight and six orders of magnitude respectively, and the network density reaches $0.611$. The \textit{US airports network} contains $N=1572$ nodes. It is also symmetric, with $w_{ij}=w_{ji}$ indicating the total passenger flow in both directions in the year 2010. The network has a density of $0.014$, much lower than that of the WTN. However, as with the latter, weights and strengths span many orders of magnitude.

We create families of filtered networks with the three schemes described above, varying the relevant parameter -- the original network is recovered for $\gamma\rightarrow 0$, $\alpha\rightarrow 1$, and $a\rightarrow 0$, respectively. For each filtered network $G^f$ with weight matrix $W^f$, we measure the ego-distance $D_{dcp}(G^f,G)$ of the filtered network from the original one, and the fraction of weight removed
\begin{equation}
	R^f=1-\frac{\sum_{ij} w^f_{ij}}{\sum_{ij} w_{ij}},
\end{equation}
i.e., the relative reduction in the total weight of $G^f$ compared to the original graph $G$.

\begin{figure*}
	\centering
	\includegraphics[scale=.5]{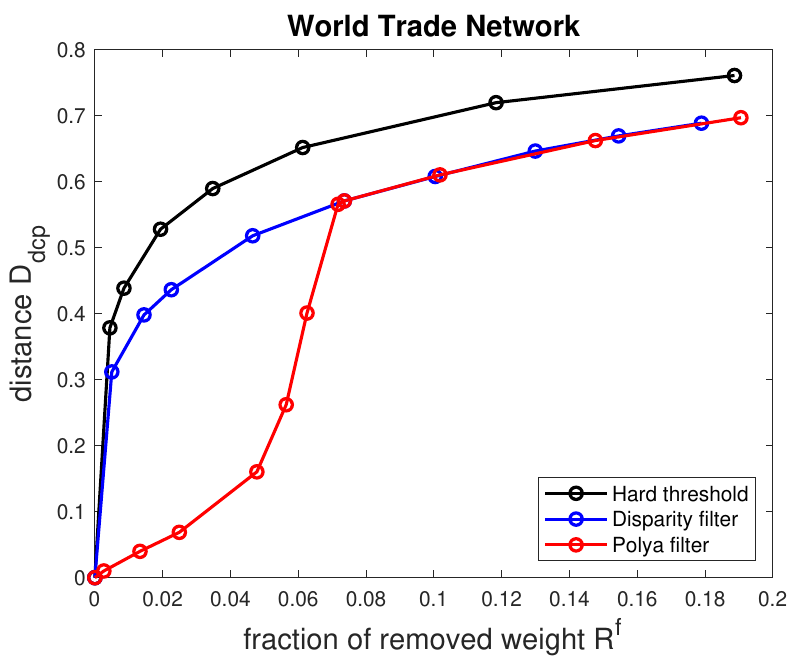}
	\includegraphics[scale=.5]{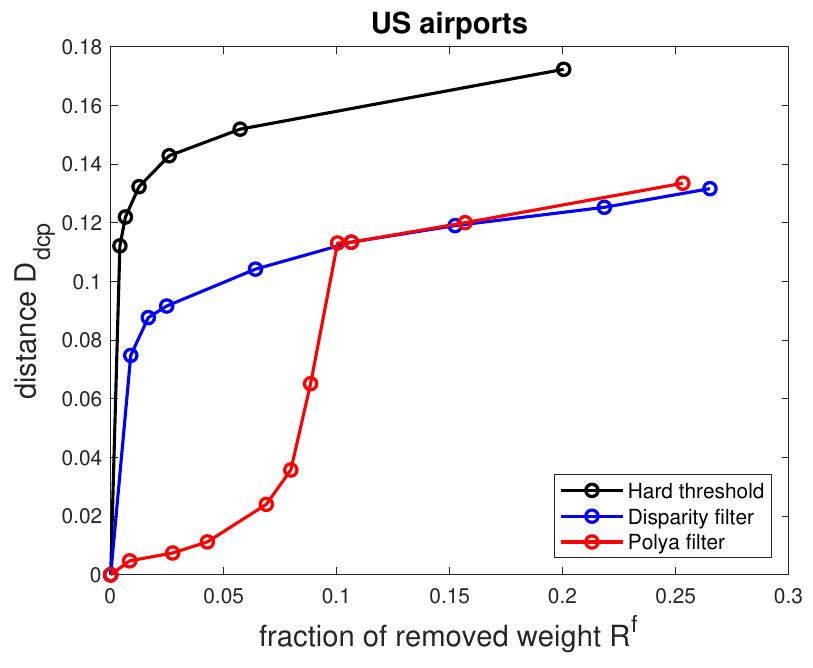}
	\caption{\label{fig:filters} Quantification of the effect of different filtering methods for multiscale weighted networks (''backbone extraction'') in terms of distance from the original network. Each dot corresponds to a network obtained by filtering the original one (WTN on the left, US airports on the right) with Hard threshold, Disparity filter, or P\'{o}lya filter (see legend). On the $x$-axis, the fraction of the total network weight removed by the filter; on the $y$-axis, the ego-distance $D_{dcp}$ from the filtered network to the original one. }
\end{figure*}

The results are summarized in Fig. \ref{fig:filters}. As expected, for both networks and all filters used, the distance increases monotonically as more and more weight is removed. However, for any given value of the removed weight, the distance from the Hard threshold graph is greater than the distance from the graphs obtained by using the nontrivial filters (Disparity and P\'{o}lya), denoting that the former tends to perturb the original  features to a larger extent. Furthermore, the P\'{o}lya filter produces backbone graphs much more similar to the original one than the Disparity filter, when the weight reduction is small. {This is consistent with the properties of the P\'{o}lya filter, which has been shown to generate more heterogeneous backbones  than those produced by other methods, i.e., more diverse than those obtained with ''naive'' thresholding\cite{Marcaccioli2019}. } We note that, at the weight reduction level $R^f$ where the two curves become almost coincident (the two methods approximately coincide when the P\'{o}lya filter parameter reaches $a=1$, as theoretically discussed\cite{Marcaccioli2019}), more than 80\% of links are removed in both study cases, while maintaining at least 90\% of the total weight. Therefore, the greater similarity with the original graph obtained from the P\'{o}lya filter is particularly appreciable, since it denotes that, for the same pruning effort, this scheme is able to preserve the egonet features of the original graph to a greater extent.

\subsection*{\label{sec:sp100}Example of application: S\&P100 financial time series}

\begin{figure*}
	\centering
	\includegraphics[scale=.6]{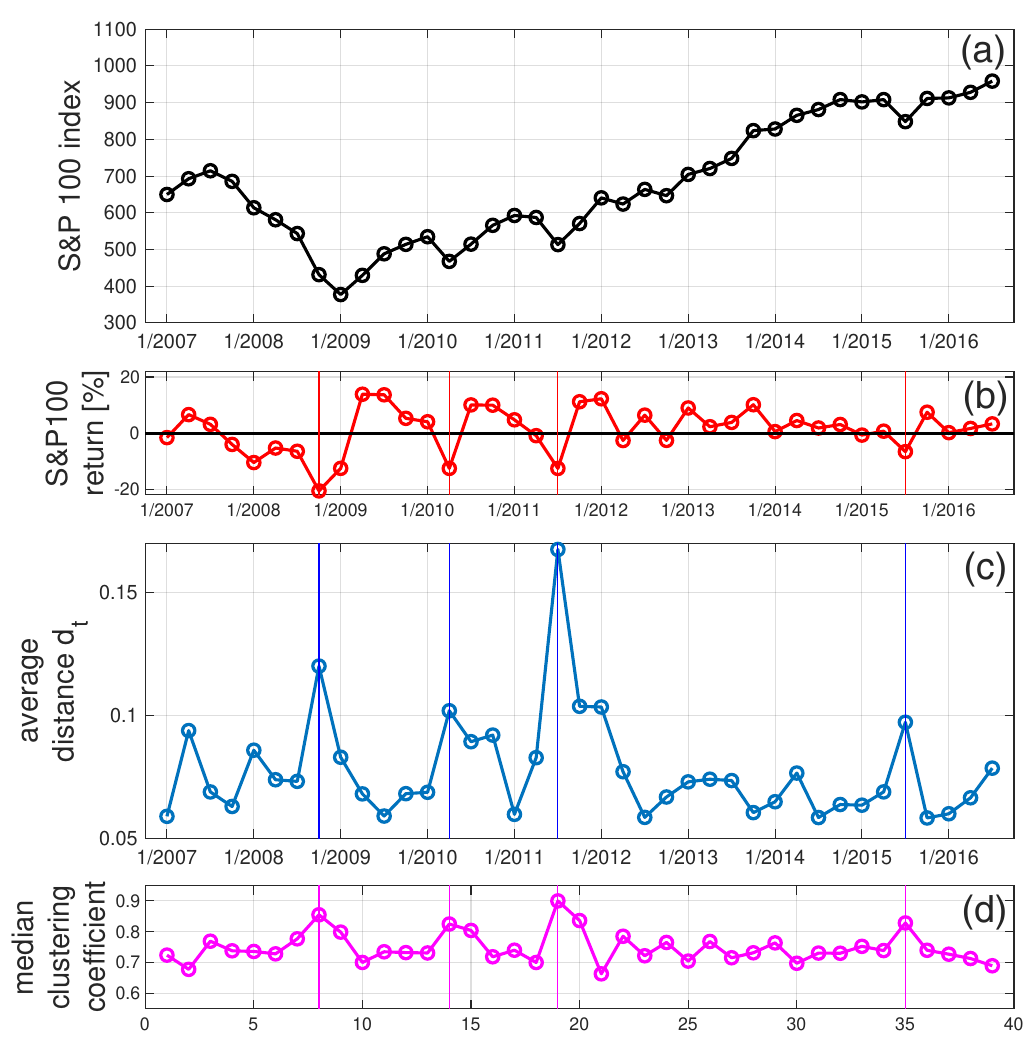}
	\caption{\label{fig:sp100} (a) Time pattern of the S\&P100 stock index in the period 2007-2016. The plot marks the latest daily value of each quarter. (b) Percentage return on a quarterly basis (c) For each quarter $t$, the plot shows the average distance $d_t$ from the graph $G^t$ to all other graphs $G^h$, $h\neq t$. Each graph $G^t$ is based on Pearson correlations between the daily time series of S\&P100 stocks in quarter $t$. (d) Median of the weighted correlation coefficient distributions }
\end{figure*}

Let us now use ego-distances to analyze data describing the stocks defining the S\&P100 index, which includes most of the largest US companies. We start from the daily closing values $y_i^\tau$ of all stocks $i$ for the 10-year period January 2007 -- December 2016 (Fig. \ref{fig:sp100}(a)) and use them to calculate the daily returns $y_i^{\tau+1}/y_i^\tau$, which are then standardized to obtain time series $u_i^\tau$ with zero mean and unit variance. The 10-year time span is then divided into quarters $t=1,2,\ldots,40$ (that is, January 2006-March 2006, April 2006-June 2006, etc.). For each quarter, a correlation network is defined whose $100$ nodes correspond to the stocks. The nodes $(i,j)$ are connected with a weight given by $w_{ij}^t=(\rho_{ij}^t+1)/2$, where $\rho_{ij}^t$ is the Pearson correlation between the daily time series $(u_i^\tau,u_j^\tau)$ for the set of days $\tau$ of the quarter $t$. Thus $w_{ij}^t$ ranges from $0$ (perfect anti-correlation) to $1$ (perfect correlation), with larger values denoting stronger agreement.

At the end of the procedure described above, we have a temporally ordered set of 40 graphs $G^t$ ($t=1,2,\ldots,40$), one for each quarter, which capture the coherence in the evolution of the stocks in each specific quarter. Note that, by construction, every graph is complete -- all links $(i,j)$ exist -- and therefore trivial in some sense. Only methods that take weights into account can be used to investigate and compare the set of graphs.

We compare the 40 graphs by computing all pairwise distances $D(G^h,G^k)$, $h,k=1,2,\ldots,40$ $(h\neq k)$. Figure \ref{fig:sp100} summarizes the results obtained using $D_{SUM}$ (see eq. (\ref{eq:sum}) -- note that, in this specific case, $D_{SUM}$ is proportional to $D_d+D_c$ since $D_p(G^h,G^k)=0$ for all $h,k$, because all graphs are complete and thus $p_i=1$ for all nodes $i$). More precisely, Fig. \ref{fig:sp100}(c) displays, as a function of $t$, the average distance $d_t$ of each graph $G^t$ from all other graphs, i.e.,
\begin{equation}\label{eq:avg_d}
	d_t=\frac{1}{39}\sum_{\substack{h=1,2,\ldots,40\\h\neq t}}D_{SUM}(G^t,G^h).
\end{equation}
The comparison of Figs. \ref{fig:sp100}(b) and (c) highlights that the four most relevant peaks (relative maxima) in the time trend of $d_t$ occur in correspondence with four large and evident (quarter) return losses. In other worlds, the correlation graph associated with financial crises has characteristics markedly different than those that characterize ordinary periods. This phenomenon has already been highlighted (e.g., Refs.\cite{ONNELA2003,Onnela2005,Lacasa2015}) and is associated with greater coherence between time series and therefore greater correlation values -- consistently, Fig. \ref{fig:sp100}(d) shows that the weighted clustering coefficients shift towards higher values.

\section*{\label{sec:conclusions}Discussion}

This paper introduced a set of dissimilarity measures to compare undirected weighted networks. The family of measures, called \textit{ego-distances}, is a generalization of those recently proposed for binary graphs\cite{Pi23} and allows one to effectively analyze all those network models in which the weights that describe the intensity of the interaction between agents are essential components of the model and cannot be ignored without unacceptable loss of information.

The proposed dissimilarity measures are based on a synthetic description of the network based on the distributions of egonet features, which capture the local structure around each node. They are alignment-free metrics, thus allowing networks to be compared even when they have different sizes (number of nodes) by focusing on their structural properties. The ability to discriminate networks with slightly different characteristics was evaluated using a standard classification setup. To overcome the lack of a suitable testbed, a new pool was defined including 12 models of weighted networks with diversified topological and weighting characteristics. Overall, we found that the ego-distances outperform the set of measures used for benchmarking and therefore should be considered as the best method currently available for all tasks requiring comparison of undirected weighted networks. Of course, this statement is based on the set of benchmark measures used for comparison, which has been defined to the best of our knowledge. Furthermore, the composition of the pool of network models obviously influences the results: despite our effort to include a large set of diversified models, both in topology and weight distribution, it must be acknowledged that a different set of models could partially alter the results.

{It is worth noting that the feature set could in principle be extended in many ways: weighted degree (strength), clustering coefficient, and persistence were chosen because they are egonet features, that is, for a node $i$ they can be computed only with knowledge of the properties of $i$ and its nearest neighbors. This simplifies the definition and implementation of the method and keeps the computational load at an acceptable level. It obviously involves a loss of information about the global graph structure, as for any embedding procedure --- but designing a network comparison method boils down to finding precisely a good trade-off between simplicity (which implies loss of information) and effectiveness. We believe that our proposal produces a satisfactory balance, as witnessed by its good overall performance. Nonetheless, the proposed scheme could naturally be extended in many ways: while keeping the node-centric flavor, one could add one or more centrality indices (e.g., betweenness) to the three egonet features. Or one could go a step further and consider two-node feature distributions, as (weighted) pairwise similarity indices. Both extensions, however, would come at the cost of much higher computational requirements.
}

As a matter of fact, a significant limitation to the widespread use of the ego-distances could be their computational cost, which scales unfavorably with network size (see \textit{Methods}) and therefore limits their applicability to small/medium sized graphs. It should be noted, however, that in principle the computation of egonet features can be fully parallelized, potentially allowing for a considerable reduction in computational requirements. Alternatively (or in combination), one could infer egonet distributions from only a sample of nodes, rather than from the entire network. This last approach, potentially promising for the analysis of large-scale networks, requires the use of appropriate graph sampling techniques\cite{Ahmed13}. We leave the in-depth evaluation of the effect of sampling on the performance of ego-distances to future research.


\section*{\label{sec:distance}Methods}

\subsection*{Egonet features}

Consider an undirected network of size (number of nodes) $N$, described by the $N\times N$ adjacency matrix $A$, with $a_{ij}=a_{ji}=1$ if $i$ and $j$ are connected by an edge, and $a_{ij}=0$ otherwise. If we denote the number of edges with $L=\frac{1}{2}\sum_{i,j=1}^{N}a_{ij}$, then the density of the network is defined as $\rho=\frac{2L}{N(N-1)}$. The network is weighted, that is, each edge $(i,j)$ is endowed with a non-negative weight $w_{ij}>0$, and the weights are collected in the $N\times N$ weight matrix $W$, with $w_{ij}=w_{ji}>0$ if $i$ and $j$ are connected by an edge, and $w_{ij}=0$ otherwise. Given node $i$, its degree $m_i=\sum_{j=1}^{N} A_{ij}$ is the number of edges incident on $i$ (i.e., number of neighbors), and its \textit{egonet} is the induced graph identified by the set of nodes $E_i=\{i\}\cup \{j|A_{ij}=1\}$, i.e., the union of node $i$ and all its neighbors, for a total of $|E_i|=m_i+1$ nodes. 

Three features, all with values in $[0,1]$, are used to characterize the egonet $E_i$: the (normalized) \textit{weighted degree} (or strength), the \textit{weighted clustering coefficient}, and the \textit{egonet persistence}. 

\begin{itemize}
	
	\item \textbf{(Normalized) weighted degree (strength).} The weighted degree $s_i=\sum_{j=1}^{N} w_{ij}$ is the sum of the weights of the edges incident on $i$. For a graph with $s_{\min}\le s_i\le s_{\max}$, we define the normalized weighted degree $d_i$ as
	\begin{equation}
		d_i=\frac{s_i-s_{\min}}{s_{\max}-s_{\min}},
	\end{equation}
	where we assume $s_{\min}\neq s_{\max}$.
	
	\item \textbf{Weighted clustering coefficient.} For an unweighted network, the clustering coefficient for a node $i$ with $m_i>1$ neighbors connected by $e_i$ edges is defined by $\tilde{c}_i=\frac{2e_i}{m_i(m_i-1)}$, while we set $\tilde{c}_i=0$ if $m_i\le 1$. Among the proposals for extension to weighted networks\cite{Barrat2004,Onnela2005,Fagiolo2007,Saramaki2007}, we adopt the one which is a generalization of the unweighted case\cite{Onnela2005} (i.e., it is reduced to the latter when all weights are equal to $1$):
	\begin{equation}\label{eq:wcc}
		c_i=\frac{2}{m_i(m_i-1)}\sum_{j,k\in E_i} \frac{(w_{ij}w_{jk}w_{ki})^{1/3}}{w_{\max}},
	\end{equation}
	where $w_{\max}=\max_{i,j=1}^{N}w_{ij}$ is the largest weight in the network and once again we set $c_i=0$ if $m_i\le 1$. As has been highlighted\cite{Onnela2005}, the weighted clustering coefficient $c_i$ can be written as the unweighted one $\tilde{c}_i$ multiplied by the \textit{average triangle intensity} at node $i$ (the intensity of a triangle is defined as the geometric mean of the weights of its three edges). Then $c_i$ combines information about the number of triangles around node $i$ and the overall weight involved in those triangles.
	
	\item \textbf{Egonet persistence.} The egonet persistence $p_i$ is defined as the persistence probability \cite{Pi11,De13} of the egonet $E_i$, i.e., the probability that a random walker, located in any of the nodes of $E_i$ at step $t$, remains at any node of $E_i$ at step $t+1$. For an undirected weighted network, it can be shown \cite{Pi11} that this quantity is equal to
	\begin{equation}\label{eq:p}
		p_i=\frac{\sum_{j\in E_i}s_j^{int}}{\sum_{j\in E_i}s_j} =\frac{\sum_{j\in E_i}s_j^{int}}{\sum_{j\in E_i}(s_j^{int}+s_j^{ext})},
	\end{equation}
	where $s_j^{int}$ (resp. $s_j^{ext}$) denotes the internal (resp. external) weighted degree of node $j$, i.e., the total weight of the edges connecting $j$ to nodes internal (resp. external) to $E_i$ (we conventionally set $p_i=0$ when $E_i=\{i\}$, i.e., $i$ is an isolated node). 
	
\end{itemize}

\begin{figure}
	\centering
	\includegraphics[width=5cm]{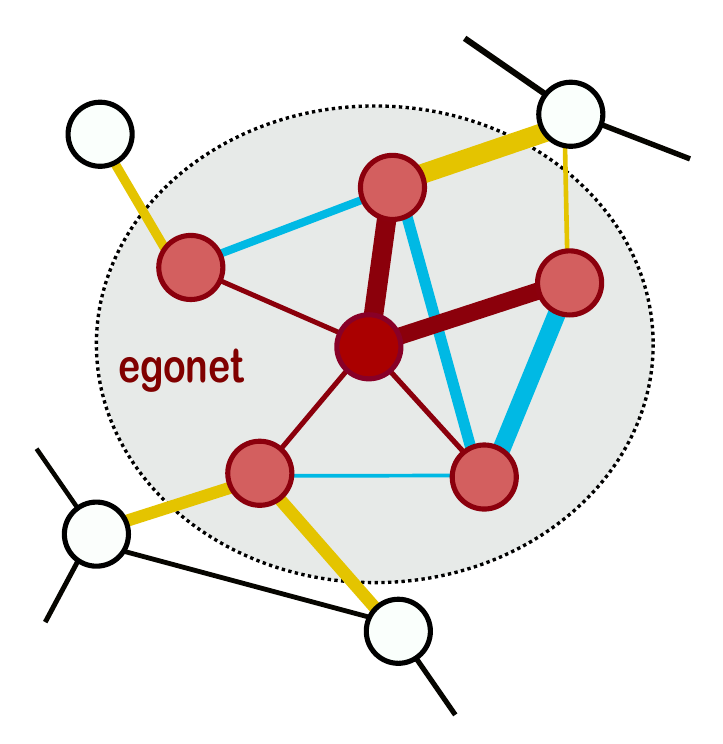}
	\caption{\label{fig:egonet} A pictorial representation of an egonet, composed of the ego-node $i$ (center) and its neighbors (red nodes): different edge thickness denotes different weight. The weighted degree $d_i$ depends only on the connectivity of node $i$ (red edges); the clustering coefficient $c_i$ also depends on the connectivity between the neighbors of $i$ (red and blue edges); the egonet persistence $p_i$ captures the balance between internal and external connectivity (red, blue, and yellow edges).}
\end{figure}

\noindent In general, the three quantities introduced above carry independent information about the egonet $E_i$ (Fig. \ref{fig:egonet}). The weighted degree $d_i$ depends only on the connectivity of node $i$ (i.e., number and weight of connections); the clustering coefficient $c_i$, however, describes the connectivity between the neighbors of $i$; finally, the egonet persistence $p_i$ captures the balance between internal and external connectivity of $E_i$, since it quantifies the proportion of strength that the egonet nodes direct into the egonet itself. 

\subsection*{\label{sec:dist}Network distances}

We calculate the \textit{weighted degree distribution} by discretizing the interval $0\le d_i \le 1$ with step $\Delta$ (so $r=1/\Delta$ is the number of intervals) and directly calculating the discrete probability distribution function (p.d.f.) $P_{d}(h)$ (i.e., the normalized histogram) by counting the proportion of ${d_i}$'s in each interval. Using the indicator function ($\mathbbm{1}_S x=1$ if $x\in S$ and zero otherwise), we can write:
\begin{equation}\label{eq:Pdd}
	P_{d}(h)=\frac{1}{N}\sum_{i=1}^{N} \mathbbm{1}_{[(h-1)\Delta,h\Delta)}{d_i},\quad h=1,2,\ldots,r,
\end{equation}
with values ${d_i}=1$ counted conventionally in the last interval $h=r$. Finally, we introduce the cumulative distribution function (c.d.f.) $Q_{d}(h)=\sum_{k=1}^{h}P_{d}(k)$, which is numerically more stable for small $N$ and therefore preferable in calculations.

In the same manner we treat the weighted clustering coefficient, i.e., we discretize the interval $0\le c_i \le 1$ and calculate the p.d.f. $P_{c}(h)$ by counting the proportion of $c_i$ in each interval, and the c.d.f. by $Q_{c}(h)=\sum_{k=1}^{h}P_{c}(k)$. We do exactly the same for the egonet persistence, obtaining the p.d.f. $P_{p}(h)$ and the corresponding c.d.f. as $Q_{p}(h)=\sum_{k=1}^{h}P_{p}(k)$.

At this point, given the two graphs $G'$ and $G''$, we define the distance between them as the (Euclidean) distance between the two corresponding c.d.f.'s $Q'_{x}$ and $Q''_{x}$, with $x\in\{d,c,p\}$ depending on whether one uses the c.d.f. of the weighted degree, the weighted clustering coefficient, or the egonet persistence, respectively. Therefore we have
\begin{equation}\label{eq:dist1}
	D_{x}(G',G'')=\frac{1}{\sqrt{r-1}}\left[\sum_{h=1}^{r}\left(Q'_{x}(h)-Q''_{x}(h)\right)^2\right]^\frac{1}{2},
\end{equation}
where the normalization term $\sqrt{r-1}$ is such that $0\le D_{x}(G',G'') \le 1$ for all $G',G''$. The above equation defines the three one-dimensional distances (i.e., based on 1-feature distributions) $D_{d}$, $D_{c}$, $D_{p}$, which are well defined -- as are the analogous ones defined below -- even when $G'$ and $G''$ have a different number of nodes, as long as the c.d.f.'s are calculated with the same step $\Delta$.

Since weighted degree, weighted clustering coefficient, and egonet persistence in general carry independent information, it is obviously possible to combine them by defining more complex metrics. A simple but effective solution is to simply aggregate them in additive form as follows:
\begin{equation}\label{eq:sum}
	D_{SUM}(G',G'')=\frac{1}{3\sqrt{r-1}}\sum_{x\in\{d,c,p\}}\left[\sum_{h=1}^{r}\left(Q'_{x}(h)-Q''_{x}(h)\right)^2\right]^\frac{1}{2}.
\end{equation} 
A more general approach is to define the two-dimensional p.d.f. $P_{xy}(h,k)$, with $x,y\in\{d,c,p\}$ $(x\neq y)$, as the normalized 2D histogram:
\begin{equation}\label{eq:P2D}
	P_{xy}(h,k)=\frac{1}{N}\sum_{i=1}^{N} (\mathbbm{1}_{[(h-1)\Delta,h\Delta)}x_i \times \mathbbm{1}_{[(k-1)\Delta,k\Delta)}y_i),\quad
	h,k=1,2,\ldots,r,
\end{equation}
with values $x_i=1$ (resp. $y_i=1$) conventionally counted in the last interval $h=r$ (resp. $k=r$). Given the two graphs $G'$ and $G''$, we then define their distance as 
\begin{equation}\label{eq:dcp}
	D_{xy}(G',G'')=\frac{1}{\sqrt{r^2-1}}\left[\sum_{h,k=1}^{r}\left(Q'_{x,y}(h,k)-Q''_{x,y}(h,k)\right)^2\right]^\frac{1}{2},
\end{equation}
that is, the (normalized) Frobenius norm of the difference between their c.d.f.'s $Q_{xy}(h,k)=\sum_{i=1}^{h}\sum_{j=1}^{k}P_{xy}(i,j)$. In this way we define three 2D measures: $D_{dc}(G',G'')$ based on the two-dimensional variable $({d_i},c_i)$; $D_{dp}(G',G'')$ based on $({d_i},p_i)$; and $D_{cp}(G',G'')$ based on $(c_i,p_i)$.

Finally, a distance measure that fully exploits all the available information is obtained by considering the multivariate distribution of the three-dimensional variable $({d_i},c_i,p_i)$. This requires partitioning the set $[0,1]^3$ into $r^3$ discretization cubes, computing the three-dimensional p.d.f. $P_{dcp}(h,k,n)$ as:
\begin{equation}\label{eq:P3D}
	P_{dcp}(h,k,n)=\frac{1}{N}\sum_{i=1}^{N}  ( \mathbbm{1}_{[(h-1)\Delta,h\Delta)}{d_i} \times  
	\mathbbm{1}_{[(k-1)\Delta,k\Delta)}c_i \times \mathbbm{1}_{[(n-1)\Delta,n\Delta)}p_i) ,\quad
	h,k,n=1,2,\ldots,r,
\end{equation}
and defining the network distance between $G'$ and $G''$ as
\begin{equation}\label{eq:d3p}
	D_{dcp}(G',G'')=\frac{1}{\sqrt{r^3-1}}
	\left[\sum_{h,k,n=1}^{r}\left(Q'_{dcp}(h,k,n)-Q''_{dcp}(h,k,n)\right)^2\right]^\frac{1}{2},
\end{equation}
where $Q_{dcp}(h,k,n)=\sum_{i=1}^{h}\sum_{j=1}^{k}\sum_{l=1}^{n}P_{dcp}(i,j,l)$ is the c.d.f.. We generically indicate by \textit{ego-distances} all the measures introduced above: $D_d$, $D_c$, $D_p$,  (1-feature distances), $D_{cp}$, $D_{dc}$, $D_{dp}$ (2-feature distances), and $D_{SUM}$, $D_{dcp}$ (3-feature distances).

\subsection*{\label{sec:models}Models of synthetic weighted networks}

We consider the set of $12$ models of undirected weighted networks described below. In all cases, a network is parameterized by the size $N$ (number of nodes) and the density $\rho={2L}/\left[N(N-1)\right]$, where $L$ is number of edges (note that the average degree $m_{avg}={2L}/{N}$ can be expressed as $m_{avg}=\rho(N-1)$). For each model, we generate networks with sizes $N=1000$, $2000$, and $4000$, and densities $\rho=0.004$, $0.01$, and $0.02$, for a total of $12\times 3\times 3=108$ combinations model/size/density. For each combination, we randomly generate $10$ network instances, so that the experimental setup includes $1080$ networks.

\begin{itemize}

\item \textbf{ER-U, ER-R, ER-D (Erd\H{o}s-R\'{e}nyi model with \textit{Uniform}, \textit{Random}, or \textit{Degree-dependent} weighting)} First, a standard ER network is created: each pair of nodes $(i,j)$, $i,j=1,2,\ldots,N$, is connected with probability $\rho$ \cite{ErRe:59,Ne:10}. Nothing else is done for a \textit{Uniform} network, which therefore has all weights equal to $1$. In the \textit{Random} case, the edge $(i,j)$ is associated with a weight $w_{ij}\sim U\left[0,1\right]$. In the \textit{Degree-dependent} case, the weight $w_{ij}=m_i m_j$ is assigned to the edge $(i,j)$.

\item \textbf{BA-U, BA-R, BA-D (Barab\'asi-Albert model with \textit{Uniform}, \textit{Random}, or \textit{Degree-dependent} weighting)} A standard BA network is first created: we define $\eta=\frac{\rho N}{2}$, which is $\frac{m_{avg}}{2}$ for large $N$ (note that $\eta$ will be integer for all $(N,\rho)$ pairs used in the article). We initialize the network with a clique (complete graph) of $\eta+1$ nodes, then add one node at a time until we reach the prescribed size $N$. Each added node must connect its $\eta$ edges to $\eta$ target nodes, which are randomly selected with probability proportional to their degree in the current network (preferential attachment \cite{BaAl:99,Ba:16}). At this point, the three weighting schemes are identical to those of ER networks (see above).

\item \textbf{GEO-U, GEO-R, GEO-D (geometric random graph model with \textit{Uniform}, \textit{Random}, or \textit{Degree-dependent} weighting)} A standard geometric random graph is first created: the $N$ nodes are thought of as points in the unit cube, whose 3D coordinates are randomly selected with uniform distribution. Then the nodes $(i,j)$ are connected if and only if their Euclidean distance is less than a given value $r>0$, whose value is iteratively adjusted to reach the prescribed density $\rho$ (on average over the 10 network replications)\cite{penrose2003}. At this point, the three weighting schemes are identical to those of ER networks (see above).

\item \textbf{YJBT (Yook-Jeong-Barab\'asi-Tu model)} The network is created following the standard BA algorithm (see above). However, when connecting a new node $j$ to a target node $i$, the edge $(j,i)$ is assigned a weight proportional to the degree $m_i$ of the target node. With the additional constraint that each new node has a fixed total weight ($=1$), we finally set $w_{ji}={m_i}/{\sum_{i'}m_{i'}}$ \cite{Yook2001}.

\item \textbf{AK-R, AK-E (Antal-Krapisvky model with \textit{Random} or \textit{Exponential} weighting)} An unweighted network is first created with a procedure similar to the BA algorithm (see above): the only difference is that, when a new node is added, it connects its $\eta$ edges to randomly selected target nodes with probability proportional to their strength (rather than their degree) in the current network (strength-driven preferential attachment\cite{Antal2005}). When the edge $(i,j)$ is added to the network, its weight is set as $w_{ij}\sim U\left[0,1\right]$ in the \textit{Random} case, and as  $w_{ij}\sim \exp(1)$ (i.e., it is drawn from an exponential distribution with mean $1$) in the \textit{Exponential} case.

\end{itemize}

\subsection*{\label{sec:classification}Classification of synthetic weighted networks}

For each pair of networks, we calculate the eight weighted ego-distances defined above (see \textit{Network distances}) using discretization step $\Delta=0.01$ (the results are largely insensitive to this parameter thanks to the use of c.d.f.'s). We also calculate the following benchmark distances\cite{Ta19}:

\begin{itemize}
	\item $D_{\text{C}_{global}}$: Based on the averaged (weighted) clustering coefficient $C=\frac{1}{N}\sum_{i=1}^{N} c_i$, the distance between two graphs with clustering coefficients $C'$ and $C''$ is  given by $D_{\text{C}_{global}}=|C'-C''|$. 
	\item $D_{SP-W/L}$ (spectral distance based on the \textit{Weight} or \textit{Laplacian} matrix): They are based, respectively, on the spectrum of the weight matrix $W$ (SP-W) or the Laplacian matrix $L=\mathrm{diag}(s_1,s_2,\ldots,s_N)-W$ (SP-L). In the SP-W case, given two networks with (symmetric) weight matrices $W'$ and $W''$ having eigenvalues $\lambda'_1\geq\lambda'_2\geq\ldots\geq\lambda'_{N_1}$ and $\lambda''_1\geq\lambda''_2\geq\ldots\geq\lambda''_{N_2}$, their distance is defined as
	\begin{equation}
		D_{SP-W}=\left[\sum_{i=1}^{N_{\min}} \left(\lambda'_i-\lambda''_i\right)^2\right]^\frac{1}{2},
	\end{equation}
	where $N_{\min}=\min\left\{N_1,N_2\right\}$. In the SP-L case, the same equation above defines $D_{SP-L}$ provided the eigenvalues of the two Laplacian matrices $L'$ and $L''$ are used \cite{Wilson2008}.
	
	\item {$D_{WD}$ (WD-metric): It is defined as a linear combination of three distance terms\cite{Jiang2021,Schieber2017}: two of them compare, with appropriate metrics, the distributions of the (weighted) shortest-path lengths in the two graphs, while the third  compares the distributions of the alpha-centralities in the two graphs and their complements.}
	
	\item $D_{PDiv}$ (Portrait Divergence): The distance is based on the comparison of the \textit{portrait matrices} $B'$ and $B''$ of the two networks, which encode the distribution of the shortest-path lengths of the graphs\cite{Bagrow2019}. We use the extension of the method to weighted networks, which requires a careful weight binning strategy described in detail in the Supplemental Material of the cited article\cite{Bagrow2019}.
	
\end{itemize}  

\noindent The evaluation of each distance is carried out within the standard Precision-Recall framework. Two networks are classified as an \textit{actual positive} pair if they originate from the same model, and as an \textit{actual negative} pair if they do not. Given a distance measure $D_x$ and a threshold $\varepsilon > 0$, a pair of networks is considered a \textit{predicted positive} sample if $D_x < \varepsilon$, and a \textit{predicted negative} sample otherwise. Precision and Recall for each $\varepsilon$ are then defined as $P_\varepsilon = TP / (TP + FP)$ and $R_\varepsilon = TP / (TP + FN)$, where \textit{TP}, \textit{FP}, and \textit{FN} represent the counts of \textit{true positive}, \textit{false positive}, and \textit{false negative} network pairs, respectively. The Precision-Recall curve visually captures the joint variation of $P$ and $R$ as $\varepsilon$ changes, and the \textit{Area Under the Precision-Recall curve} (AUPR), with $0 \leq \text{AUPR} \leq 1$, encapsulates the performance of the distance measure. The ideal case corresponds to an AUPR of 1\cite{Davis2006,Saito2015}.

\subsection*{\label{sec:computation}Computational requirements}

The computation of the ego-distances defined above requires a sequence of operations of a mixed nature, i.e., the calculation of one or more egonet features for each single node, followed by the calculation of the c.d.f.'s for each graph and finally of their distance. A theoretical prediction of the computational requirements is not simple, although the first task (computing egonet features) is certainly dominant for medium to large networks.

Consider a network with $N$ nodes. The time required to read the weight $w_{ij}$ of a connection $(i,j)$ is constant, meaning that computing the weighted degree of a single node takes $O(N)$ time, and for all nodes, this results in $O(N^2)$ time. For calculating the clustering coefficient (eq. \eqref{eq:wcc}), determining the weights of all connections between neighbors of a node with degree $m_i$ requires $O(m_i^2)$ operations. In the worst case, this is $O(N^2)$, and for the entire network, the complexity is $O(N^3)$. Similarly, for the egonet persistence (eq. \eqref{eq:p}), the numerator involves reading the weights of the $(m_i + 1)^2$ possible internal connections to $E_i$, while the denominator requires reading the $(m_i + 1)N$ connections between the nodes of $E_i$ and all other nodes in the network. In the worst case, both terms are $O(N^2)$, resulting in an overall complexity of $O(N^3)$ for the entire network. Typical networks are often sparse and therefore quite far from the worst case: we expect lighter computational requirements, but still in the range $O(N^2)$ to $O(N^3)$. Indeed, empirical analyzes on the same heterogeneous set of synthetic networks described above obtain a computation time approximately increasing as $t\propto N^\alpha$, with $\alpha$ ranging between $2.32$ and $2.50$ for the distances which uses all the three egonet features (see \textit{Supplementary Information} for details). On the other hand, the worst-case computational requirement of the benchmark distances used for comparison varies from $O(LN+N^2\log N)$ for Portrait Divergence on weighted networks\cite{Bagrow2019} (which is $O(N^3)$ in the worst case $L\approx N^2$ and $O(N^2\log N)$ for sparse networks $L\approx N$), to $O(N^3)$ for spectral methods that require the computation of $N$ eigenvalues\cite{Pan1999}.

\subsection*{Code and data availability}
The Matlab code of the function EgoDistW, which implements the computation of ego-distances, and the files of the synthetic networks used for the classification task are available at \url{https://piccardi.faculty.polimi.it/highlights.html}.\\
WTN data is available at \url{http://www.cepii.fr/CEPII/en/bdd_modele/bdd_modele.asp}. US airports data was downloaded from \url{https://toreopsahl.com/datasets}. 
The official webpage of the S\&P100 stock index is \url{https://www.spglobal.com/spdji/en/indices/equity/sp-100/#overview}. 



\section*{Acknowledgements}
C.P. was partially supported by the project “CODE – Coupling Opinion Dynamics with Epidemics”, funded under PNRR Mission 4 "Education and Research" - Component C2 - Investment 1.1 - Next Generation EU "Fund for National Research Program and Projects of Significant National Interest" PRIN 2022 PNRR, grant code P2022AKRZ9, CUP B53D23026080001. The author is grateful to the two anonymous Reviewers for their helpful and constructive comments.

\section*{Author contributions statement}
C.P. conceived the research, conducted the experiments, and wrote the manuscript. 

\section*{Additional information}
The author declares no competing interest.

\end{document}


\maketitle



\section*{\label{sec:size_density}{Size/density effects}}

{In the main paper (see \textit{Results/Model Classification}) we discussed the classification performance of the proposed ego-distances and of the benchmark distances in two scenarios, namely when only networks with the same size and density are compared and, on the other hand, when size and density are mixed and act as confounders in recognizing whether two graphs come from the same generating model.
	
Here we illustrate additional experiments aimed at studying the effects of graph size $N$ and density $\rho$ on the classification performance separately. For brevity, we report only the results of the two 3-feature ego-distances ($D_{SUM}$ and $D_{dcp}$), which proved to achieve the best average performance, and compare them with three benchmark measures ($D_{\text{C}_{global}}$, $D_{WD}$ and $D_{PDiv}$). The results are summarized in Supplementary Table \ref{tab:breakdown}.
	
From examining the table, we first note that for any combination of size and density, ego-distances always outperform the benchmark measures --- and $D_{dcp}$ is always dominant over $D_{SUM}$, albeit moderately (this is consistent with the general results discussed in the main paper). Second, the performance of all five measures monotonically improves by increasing $N$ (upper half of the table) or by increasing $\rho$ (lower half). With different approaches, all measures are effectively based on the computation of appropriate graph summary statistics: it is clear that classification performance benefits from increasing the available information.

}

\begin{table}
	\centering
	\caption{\label{tab:breakdown} {AUPR (Area Under the Precision/Recall curve) value for the classification of network models, for the two 3-feature ego-distances $D_{SUM}$ and $D_{dcp}$ and for three benchmark distances. The classification is performed separately for networks of the same size in the upper half of the table, including all densities; and for networks of the same density in the lower half, including all sizes. } }
	{
		\begin{tabular}{c|ccc}
			\multicolumn{1}{c}{} & \multicolumn{1}{|c}{$N=1000$, all densities} & \multicolumn{1}{c}{$N=2000$, all densities} & \multicolumn{1}{c}{$N=4000$, all densities} \\
			\toprule
			$D_{SUM}$ & 0.510 & 0.544 & 0.554 \\
			$D_{dcp}$ & 0.567 & 0.602 & 0.606 \\
			\midrule
			$D_{\text{C}_{global}}$ & 0.233 & 0.246 & 0.273 \\		
			$D_{WD}$ & 0.300 & 0.340 & 0.356 \\	
			$D_{PDiv}$ & 0.408 & 0.468 & 0.506 \\	
			\bottomrule		
			\multicolumn{1}{c}{} & \multicolumn{1}{|c}{$\rho=0.004$, all sizes} & \multicolumn{1}{c}{$\rho=0.01$, all sizes} & \multicolumn{1}{c}{$\rho=0.02$, all sizes} \\
			\toprule
			$D_{SUM}$ & 0.481 & 0.581 & 0.691 \\
			$D_{dcp}$ & 0.514 & 0.623 & 0.733 \\
			\midrule
			$D_{\text{C}_{global}}$ & 0.309 & 0.493 & 0.616 \\		
			$D_{WD}$ & 0.332 & 0.370 & 0.393 \\	
			$D_{PDiv}$ & 0.428 & 0.471 & 0.505 \\	
			\bottomrule		
	\end{tabular} }
\end{table}


\clearpage

\section*{\label{sec:computation}Computational requirements}

To obtain empirical confirmation of the estimated computational requirements of the ego-distances (see \textit{Methods/Computational requirements} in the main paper), we leverage the same pool of synthetic networks previously used to evaluate the classification capabilities of the proposed ego-distances. We perform experiments for all nine combinations of size ($N=1000$, $2000$, and $4000$) and density ($\rho=0.004$, $0.01$, and $0.02$) used above: for each pair $(N,\rho)$, we compute the $120\times 119/2=7140$ distances between all possible pairs of the $120$ networks having the prescribed size and density (recall that we have $12$ network models and $10$ replicas for each model), and we add up the time needed for the computation. Finally, we obtain an aggregate time for each pair $(N,\rho)$, which depends on models with mixed characteristics and is therefore representative of the average computational needs of the distance used.

\begin{figure*}[h]
	\centering
	\includegraphics[width=17.5cm]{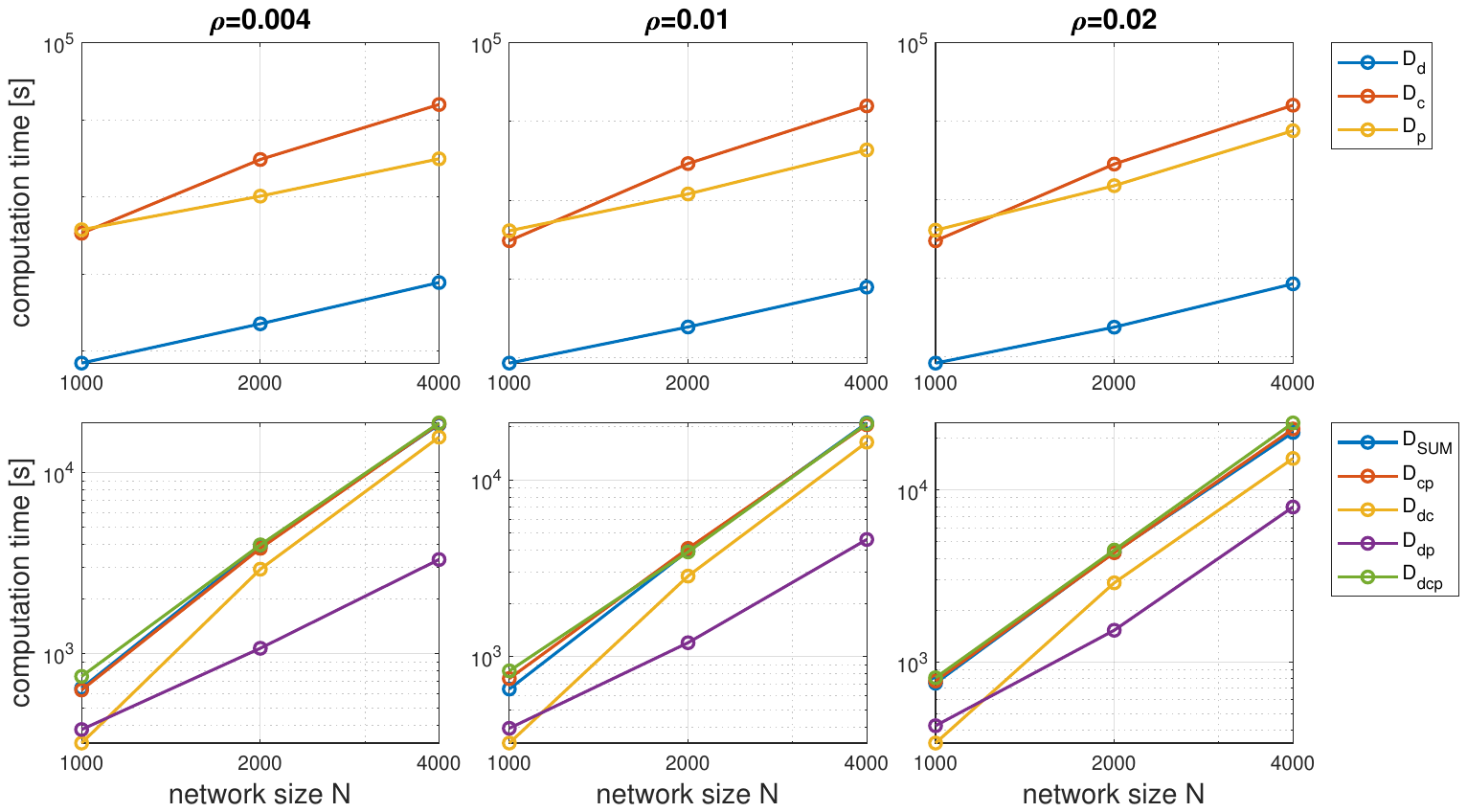}
	\caption{\label{fig:time_N_rho} Computation time as a function of the network size $N$ for different densities $\rho$, for the ego-distances defined in the main paper (see \textit{Methods}), $\Delta=0.01$. Each point is the aggregated time of computing the $120\times 119/2=7140$ distances between all possible pairs of the $120$ networks having the prescribed size and density ($12$ network models $\times$ $10$ replications). In the lower panels, the curve $D_{SUM}$ is practically invisible because it almost coincides with the curve $D_{cp}$.  }
\end{figure*}

Supplementary Figure \ref{fig:time_N_rho} shows the results of the experiments described above. First, regarding 1-feature distances, we note that, as expected, the time needed to compute $D_d$ is much shorter than that needed for $D_p$ and $D_c$, the latter being the most demanding (top panels). Consequently, when considering 2- and 3-feature distances (bottom panels), those that include both $c$ and $p$ require the highest computational effort. Obviously $D_{dcp}$, which exploits the three egonet features in the most structured form, requires the greatest effort in all cases, even if with a very small gap compared to $D_{cp}$ and $D_{SUM}$. 

All panels in Supplementary Figure \ref{fig:time_N_rho} show, for all $\rho$ values, a computation time that increases approximately as $t\propto N^\alpha$, with $\alpha$ ranging between $1.53$ and $2.87$ (from $2.32$ to $2.50$ if we only consider 3-feature distances). However, the comparison between the columns of Supplementary Figure \ref{fig:time_N_rho} reveals that the computation time is almost independent from the density $\rho$.

The computation times shown in Supplementary Figure \ref{fig:time_N_rho} were obtained on a desktop PC with Intel i7 CPU at 2.90GHz using Matlab R2021b. To limit possible confounding factors, the  times reported refer only  to the computation of distances between the $7140$ network pairs, as described above, while all data loading and organization are ignored. We used the code EgoDistW available at \url{https://piccardi.faculty.polimi.it/highlights.html}, which implements all the distances proposed in this paper. 

